\documentclass[referee]{raa} % referee version: for submission manuscript
\usepackage{graphicx,times}
\usepackage{natbib}
\usepackage{amssymb,amsmath}
\bibpunct{(}{)}{;}{a}{}{,}

\renewcommand{\vec}[1]{\mbox{\boldmath $#1$}}

\usepackage[pagebackref=true]{hyperref}
\hypersetup{colorlinks = true, linkcolor=red, citecolor=blue, filecolor=cyan, urlcolor=magenta}

\begin{document}

  \title{
  The Dependence of Stellar Activity Cycles on Effective Temperature
  }
%   \subtitle{I. Place Your Subtitle Here}

   \volnopage{Vol.0 (20xx) No.0, 000--000}      %%preserved for Editor. DOn't remove!
   \setcounter{page}{1}          %%starting page, preserved for Editor. DOn't remove!

   \author{Leonid Kitchatinov}

   \institute{Institute of Solar-Terrestrial Physics SB RAS, Lermontov Str. 126A, 664033, Irkutsk, Russia; {\it kit@iszf.irk.ru}\\
\vs\no
   {\small Received 2022 May 13; accepted 2022 September 30}}

%%%%%%%%%%%%%%%%%%%%%%%%%%%%%%%%%%%%%%%%%%%%%%%%%%%%%%%%%%%%%%%%%%%%%%%%%%%%%%%%%%%%%%%
\abstract{This paper proposes the idea that the observed dependence of
stellar activity cycles on rotation rate can be a manifestation of a
stronger dependence on the effective temperature. Observational evidence
is recalled and theoretical arguments are given for the presence of cyclic activity in the case of sufficiently slow rotation only. Slow rotation means proximity to the observed upper bound on the rotation period of solar-type stars. This maximum rotation period depends on temperature and shortens for hotter stars. The maximum rotation period
is interpreted as the minimum rotation rate for operation of a
large-scale dynamo. A combined model for differential rotation and the
dynamo is applied to stars of different mass rotating with a rate
slightly above the threshold rate for the dynamo. Computations show
shorter dynamo cycles for hotter stars.
As the hotter stars rotate faster, the computed cycles are also shorter for faster rotation. The observed smaller upper bound for rotation period of hotter stars can be explained by the larger threshold amplitude of the $\alpha$-effect for onset of their dynamos: a larger $\alpha$ demands faster rotation.
The amplitude of the (cycling) magnetic energy in the computations is proportional to the difference between the rotation period and its upper bound for the dynamo. Stars with moderately different rotation rates can differ significantly in
super-criticality of their dynamos and therefore in their magnetic
activity, as observed.
%%%%%%%%%%%%%%%%%%%%%%%%%%%%%%%%%%%%%%%%%%%%%%%%%%%%%%%%%%%%%%%%%%%%%%%%%%%%%%%%%%%%%%%%
}
%%%%%%%%%%%%%%%%%%%%%%%%%%%%%%%%%%%%%%%%%%%%%%%%%%%%%%%%%%%%%%%%%%%%%%%%%%%%%%%%%%%%%%%%

   \authorrunning{L.\,L.~Kitchatinov}            %author_head in even pages
   \titlerunning{Temperature Dependence of Stellar Cycles}  % title_head in odd pages

   \maketitle

%%%%%%%%%%%%%%%%%%%%%%%%%%%%%%%%%%%%%%%%%%%%%%%%%%%%%%%%%%%%%%%%%%%%%%%%%
\section{Introduction}\label{S1}
%%%%%%%%%%%%%%%%%%%%%%%%%%%%%%%%%%%%%%%%%%%%%%%%%%%%%%%%%%%%%%%%%%%%%%%%%
Large-scale magnetic fields and large-scale flows in late-type stars are believed to be caused by global rotation. Accordingly, many observational and theoretical studies are focused on
the dependence of stellar magnetic activity and/or differential rotation
on the rotation rate. This paper suggests that magnetic activity
{\em cycles} and differential rotation are more sensitive to another
stellar parameter of the effective temperature.

The temperature dependence can manifest itself as seeming dependence on rotation rate because of the following. Solar-type stars are spinning-down with age due to the angular momentum loss for magnetically coupled wind \citep{Kraft_67}.
Proportionality constant in the \citet{Skumanich_72}
law $P_{\rm rot} \propto t^{1/2}$ for stellar spindown is a decreasing function of temperature \citep{Barnes_07}.
Among stars of approximately the same age $t$, cooler stars have longer rotation period $P_{\rm rot}$. Relatively fast rotation in a sample of (solar-type) stars is usually represented by F-stars while cooler K-stars are on the slow rotation side of the sample \citep[see fig.\,3 in][as a characteristic example]{Donahue_EA_96}.
A rotation rate dependence inferred from the sample can therefore include an implicit   dependence on temperature.

In the case of differential rotation, this statement is more or less
evident by now. Observations \citep{Don_Cam_97,Barnes_EA_05,Bal_Abedi_16}
and theoretical modelling \citep{Kit_Rue_99,Kit_Ole_12DR}
both suggest that former detections of an increasing trend in dependence of the differential rotation on rotation rate can result from combining a strong increase of the differential rotation with temperature with its moderate dependence on rotation rate.

This paper considers the possibility that stellar activity cycles can
also be more dependent on temperature than on the rate of rotation.
Observations of stellar cycles were mainly focused on the dependence on
rotation rate. A tendency of shorter cycles for faster rotation has been
found \citep{Noyes_EA_84} though with a considerable scatter and
possible discontinuities in this trend
\citep{Brandenburg_Cyc_98,Brandenburg_Cyc_99,Bom-Vitense07Two_branches}.
Several attempts at reproducing the trend with dynamo models were
undertaken with mixed success
\citep{Brun_Cyc_10,Karak_EA_14,Hazra_EA_19,Pipin_Cyc_21}. There is
an extensive literature on direct numerical simulations of global
stellar convection including simulations of the activity cycles
\citep[see][and references therein]{Strugarek_EA_17_3D_Cycles,
Warnecke_18_3D_Cycles, Strugarek_EA_18_3D_Cycles, Brun_EA_22_3D_Cycles}.
The decreasing trend in cycle duration with rotation rate remains
however unexplained.

Observations of stellar rotation revealed the upper bound on rotation
period of solar-type stars \citep{Rengarajan_84,van_Saders_EA_16}.
Magnetic activity is low near this maximum rotation period
\citep{Metcalfe_EA_16}. The maximum period can therefore be interpreted
as a minimum rotation rate for large-scale stellar dynamos
\citep{Cameron_Schussler_17,Kit_Nep_17,Metcalfe_Egeland_19}. We keep to
this interpretation and apply a joint mean-field model for differential
rotation and dynamo to stars of different mass rotating with a period
close to its observationally detected upper bound. The computations show
an increase in differential rotation and shortening of the dynamo-cycle
for stars of higher mass and effective temperature. As the maximum
rotation period is shorter for hotter stars \citep{van_Saders_EA_19},
the temperature trends also imply shorter cycles and larger differential
rotation for faster rotation, though the rotation rate is not the
main physical parameter governing the trends.

We also recall observational evidence and discuss theoretical arguments
for the presence of activity cycles in sufficiently slow rotating stars
only. This is done in the following Section~\ref{S2}. Section~\ref{S3}
explains our method and the dynamo model. Section~\ref{S4} presents and
discusses the results. Section~\ref{S5} summarises the results and
concludes.
%%%%%%%%%%%%%%%%%%%%%%%%%%%%%%%%%%%%%%%%%%%%%%%%%%%%%%%%%%%%%%%%%%%%%%%%%
\section{Should cyclic activity be expected for rapid rotators?}\label{S2}
%%%%%%%%%%%%%%%%%%%%%%%%%%%%%%%%%%%%%%%%%%%%%%%%%%%%%%%%%%%%%%%%%%%%%%%%%
The project of long-term monitoring of chromospheric activity at the
Mount Wilson Observatory revealed activity cycles similar to the 11-year
solar cycle on many sun-like stars \citep{Wilson_78,Baliunas_EA_95}.
Summarising the results of the project, \citet{Baliunas_EA_95} noted
that cycles on young rapid rotators are rare but slow rotators as old as
the Sun have cycles (except for the cases of flat and low activity like
the solar Maunder minimum). Recently, \citet{Chrom_Act_Cat_18} compiled
a chromospheric activity catalog of Mount Wilson data and more recent data on
solar-type stars. Their table\,4 gives stars with cyclic activity.

\begin{figure}
\centering
\includegraphics[width= 12 truecm]{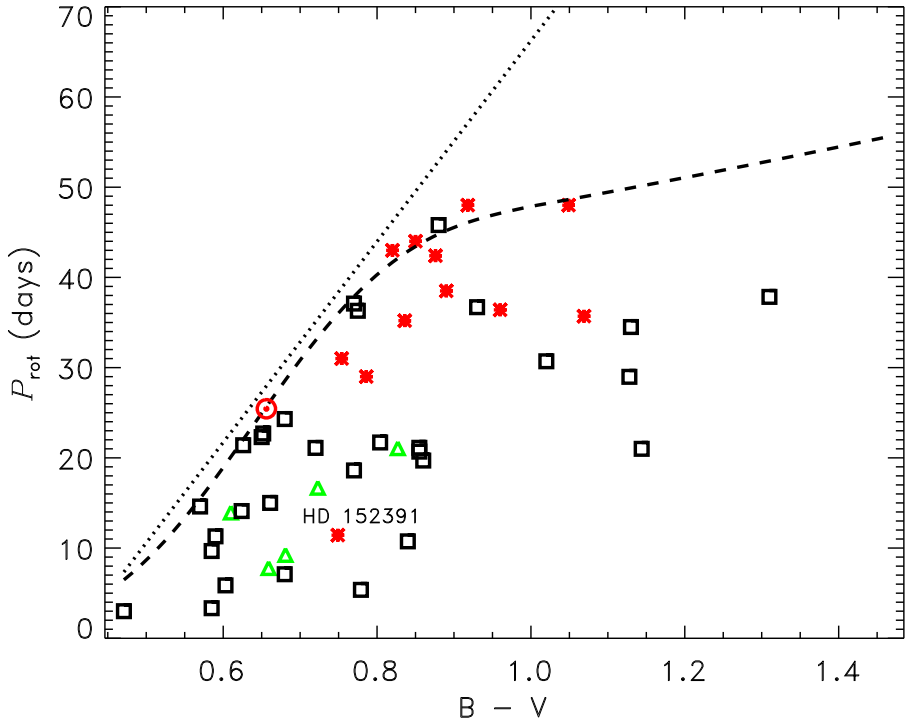}
   \caption{Positions of the main-sequence stars showing activity cycles on the
            $P_{\rm rot}-(B - V)$ plane. Red symbols show stars with well-defined solar-like activity cycles. Green triangles show stars with multiple cycles and black squares - the stars with uncertain \lq\lq probable" cycles. The dotted and dashed lines show the maximum rotation period $P_{\rm max}$ for stellar spindown according to  \citet{Rengarajan_84} and \citet{van_Saders_EA_19} respectively.}
\label{f1}
\end{figure}

Figure~\ref{f1} shows positions of the stars with activity cycles from
the catalog by \citet{Chrom_Act_Cat_18} on the plane of the rotation
period $P_{\rm rot}$ and the $B-V$ colour. Only main-sequence stars are
included in the plot\footnote{Some inaccuracies were corrected in
table\,4 by \citet{Chrom_Act_Cat_18} when producing Fig.\,\ref{f1}: The
color $B - V = 0.594$ given in the table for HD\,160346 is too small for
the K3 star. It was corrected to the value of 0.971 observed
\citep{Hog_EA_00}. The HD\,81809 misclassified in the table as a
main-sequence star is a binary system whose active component is subgiant
\citep{Egeland_18}. Subgiants are not included in Fig.\,\ref{f1}.}.

Figure\,\ref{f1} shows also the observed upper bound $P_{\rm max}$ on the rotation period. The dotted line in Fig.\,\ref{f1} is the linear approximation for the $P_{\rm max} -
(B-V)$ relation by \citet{Rengarajan_84}. Rengarajan found this
approximation for $B - V < 1$. Based on a vast statistics of recent data
on stellar rotation, \citet{van_Saders_EA_19} found that the $P_{\rm
max} - (B-V)$ relation is well approximated by the constant value
\begin{equation}
    {\rm Ro} = 2.08
    \label{1}
\end{equation}
of the Rossby number ${\rm Ro} = P_{\rm rot}/\tau_{\rm c}$; the
convective turnover time $\tau_{\rm c}$ is a function of $B-V$ colour
\citep[see eq.\,(4) in][]{Noyes_EA_84_1}. The $P_{\rm max}$ after
Eq.\,(\ref{1}) is shown by the dashed line in Fig.\,\ref{f1}. Proximity
to this line can quantify the meaning of \lq slow rotation' for the
main-sequence dwarfs.

Following \citet{Chrom_Act_Cat_18}, the quality of the activity cycles in Fig.\,\ref{f1}
is encoded by colour: red symbols show well defined
solar-like cycles, green triangles show stars with multiple cycles, and
black squares - stars with \lq\lq probable" cycles.
Almost all stars with well-defined cycles are slow rotators with $P_{\rm rot}$ close to $P_{\rm max}$. The only exception is HD\,152391.
\citet{Olspert_EA_18} found doubly periodic activity for this star with their harmonic
regression model. This may explain the position of HD\,152391 in the
region of green symbols in Fig.\,\ref{f1}.

Also on physical grounds, it seems plausible to expect cyclic activity
for slow rotators only.

Hydromagnetic dynamos of any kind can be understood as the {\em
instability} of conducting fluids to magnetic disturbances: only if
whatever small but finite seed field is present, can a
dynamo-instability amplify and support the field. Similar to all other
instabilities, dynamo-instability has dimensionless governing parameters
and onset when the parameters exceed a certain threshold value. The
empirical Eq.\,(\ref{1}) can be seen as the normalised rotation rate for
the onset of large-scale stellar dynamos. The red symbols of
Fig.\,\ref{f1} close to the $P_{\rm max}$-line represent slightly
supercritical cyclic dynamos.

Various instabilities behave similarly in dependence
on their governing parameters. For slightly supercritical parameters,
new steady (change of stability) or oscillatory (overstability) states
are normally realised \citep{Chandrasekhar_61}. In a highly
supercritical case, instabilities change - as a rule - to a turbulent
regime possibly with an intermediate stage of multi-periodic dynamics
\citep[see, e.g.,][]{Landafshitz_Hydro_87}. If large-scale dynamo
instability is not an exception to this rule, some kind of dynamo
turbulence with non-cyclic activity should be expected for the highly
supercritical regime of rapid rotation.

The expectation  is hard to test with computations. This would require
fully nonlinear and highly supercritical dynamo models lacking at the
moment. The strongly nonlinear regime of dynamos in rapid rotators is in
particular indicated by their observed large torsional oscillations
\citep{Cameron_Donati_02}. Changes to turbulence usually proceed via
break of symmetry of slightly supercritical regimes. An adequate dynamo
model has to be nonlinear and non-axisymmetric.

Lacking such a dynamo model, computations of this paper are restricted to slightly
supercritical cyclic dynamos of stars with rotation period close to $P_{\rm max}$ of
Fig.\,\ref{f1} corresponding to the constant Rossby number of Eq.\,(\ref{1}).
%%%%%%%%%%%%%%%%%%%%%%%%%%%%%%%%%%%%%%%%%%%%%%%%%%%%%%%%%%%%%%%%%%%%%%%%%
\section{Model and method}\label{S3}
%%%%%%%%%%%%%%%%%%%%%%%%%%%%%%%%%%%%%%%%%%%%%%%%%%%%%%%%%%%%%%%%%%%%%%%%%
\subsection{Combined model of differential rotation and dynamo}\label{S3.1}
%%%%%%%%%%%%%%%%%%%%%%%%%%%%%%%%%%%%%%%%%%%%%%%%%%%%%%%%%%%%%%%%%%%%%%%%%
We apply a joint model of dynamo and differential rotation by
\citet{Kit_Nep_17AL,Kit_Nep_17} to stars of different mass $0.7 \leq
M/M_\odot \leq 1.2$ and rotation period close to $P_{\rm max}$ of
Eq.\,(\ref{1}). Differential rotation and meridional flow for dynamo
computations are supplied by an axisymmetric hydrodynamical mean-field
model. The model differs from that of \citet{Kit_Ole_11DR} only in a
modification of the mixing length $\ell$: the mixing length of its
standard definition $\ell_0 = \alpha_{MLT}H_{\rm p}$ ($H_{\rm p}$ is the
pressure scale height) is now reduced near the inner boundary $r_{\rm
i}$ of the convection zone so that it can exceed the distance to the
boundary only slightly:
\begin{equation}
    \ell = \ell_\mathrm{min} +\frac{1}{2}\left(\ell_0 - \ell_\mathrm{min}\right)
    \left[ 1 + \mathrm{erf}\left(\frac{r/r_\mathrm{i} - x_\ell}{d}\right)\right] .
    \label{2}
\end{equation}
In this equation, $\ell_{\rm min} = 0.01 R$ equals one percent of the stellar radius, $\mathrm{erf}$ is the error function, and other parameters will be specified later. Our differential rotation model differs from other mean-field formulations in that it does not prescribe the eddy transport coefficients but computes them. The eddy viscosity in particular is defined by the equation
\begin{equation}
    \nu_{_\mathrm{T}} = -\frac{\tau \ell g}{15 c_{\rm p}}\frac{\partial S}{\partial r} ,
    \label{3}
\end{equation}
where $g$ is gravity, $\tau$ is the (position dependent) convective
turnover time, $c_{\rm p}$ is the specific heat capacity at constant
pressure. The specific entropy $S$ in Eq.\,(\ref{3}) is a dependent
variable of the model. The entropy is controlled by the (nonlinear) heat
transport equation that is one of three equations of the model (the
other two being the equations for the meridional flow and angular
velocity). Recently \citet{Jermin_EA_18} discussed the performance of
this closure method in the convective turbulence theory.

We avoid repeating other details of the differential rotation model all
of which can be found elsewhere \citep{Kit_Ole_11DR,Kit_Ole_12DR}.

Our dynamo model is a particular version of the flux-transport models
pioneered by \citet{Choudhury_EA_95} and \citet{Durney_95}. The models'
name reflects the importance of magnetic field advection by the
meridional flow. The flux-transport models with the $\alpha$-effect of
Babcock-Leighton (BL) type agree closely with solar observations
\citep{Jiang_EA_13,Charbonneau_20}.

Our dynamo model is formulated for a spherical layer of a stellar
convection zone. The standard spherical coordinate system
($r,\theta,\phi$) with the rotation axis as the polar axis is used. The
formulation assumes axial symmetry of the mean magnetic field
\begin{equation}
    {\vec B} = \hat{\vec\phi} B + {\vec\nabla}\times
    \left(\hat{\vec\phi}\frac{A}{r\sin\theta}\right)
    \label{4}
\end{equation}
and flow
\begin{equation}
    {\vec V} = \hat{\vec\phi}\, r\sin\theta\,\Omega + \rho^{-1}{\vec\nabla}\times
    \left(\hat{\vec\phi}\frac{\psi}{r\sin\theta}\right).
    \label{5}
\end{equation}
In these equations, $B$ is the toroidal magnetic field, $A$ is the poloidal field potential, $\Omega$ is the angular velocity, $\psi$ is the stream function for the meridional flow, $\hat{\vec\phi}$ is the azimuthal unit vector, and $\rho$ is density.

Two joint dynamo equations for the poloidal and toroidal magnetic fields read
\begin{eqnarray}
    \frac{\partial A}{\partial t} &=& \frac{1}{\rho r^2\,\sin\theta}
    \left(\frac{\partial\psi}{\partial r}\frac{\partial A}{\partial\theta}
    - \frac{\partial\psi}{\partial\theta}\frac{\partial A}{\partial r}\right)
    + r\sin\theta\,{\cal E}_\phi\ ,
    \label{6} \\
    \frac{\partial B}{\partial t} &=&
    \frac{1}{\rho r^2}\frac{\partial\psi}{\partial r}
    \frac{\partial}{\partial\theta}\left(\frac{B}{\sin\theta}\right)
    - \frac{1}{r\sin\theta}\frac{\partial\psi}{\partial\theta}
    \frac{\partial}{\partial r}\left(\frac{B}{\rho r}\right)
    \nonumber \\
    &+& \frac{1}{r}\left(\frac{\partial\Omega}
    {\partial r}\frac{\partial A}{\partial\theta}
    - \frac{\partial\Omega}{\partial\theta}\frac{\partial A}{\partial r}
    + \frac{\partial (r {\cal E}_\theta)}{\partial r}
    - \frac{\partial{\cal E}_r}{\partial\theta}\right)\ ,
    \label{7}
\end{eqnarray}
where ${\vec{\cal E}} = \langle {\vec u}\times{\vec b}\rangle$ is the mean electromotive force \citep[EMF,][]{Krause_Raedler_80}, which results from a correlated action of fluctuating velocities $\vec u$ and magnetic fields $\vec b$ and includes all the dynamo-relevant  effects of convective turbulence.

Expression for the EMF can be rather complicated \citep{Pipin_08EMF}.
Some simplification can be achieved by splitting the EMF in three parts,
\begin{equation}
    {\vec{\cal E}} = {\vec{\cal E}}^\alpha + {\vec{\cal E}}^{\rm diff}
    + {\vec{\cal E}}^{\rm dia},
    \label{8}
\end{equation}
responsible for the $\alpha$-effect, eddy diffusion, and diamagnetic
pumping respectively.

Toroidal field generation by the $\alpha$-effect is neglected in the
$\alpha\Omega$-dynamo. Nonlocal $\alpha$-effect of BL type is prescribed
in the poloidal field equation (\ref{6})
\begin{equation}
    {\cal E}^\alpha_\phi =
    \frac{\alpha\,B(r_{\rm i},\theta)}{1 + (B(r_{\rm i},\theta)/B_0)^2}
    \cos\theta\sin^{n_\alpha}\theta\,\phi_\alpha(r/r_{\rm e}),
    \label{9}
\end{equation}
where the function
\begin{equation}
    \phi_\alpha(r/r_{\rm e}) = \frac{1}{2}\left[ 1 +
    {\rm erf}\left((r/r_{\rm e} + 2.5h_\alpha - 1)/h_\alpha\right)\right],
    \label{10}
\end{equation}
with $h_\alpha = 0.02$ peaks near the external boundary $r_{\rm e} =
0.97R$. This boundary is placed shortly below the stellar surface to
exclude the near-surface layer of steep stratification that is difficult
to include in the differential rotation model. The $\alpha$-effect of
Eq.\,(\ref{9}) describes generation of the poloidal field near the
surface from the bottom toroidal field. The large value of $n_\alpha =
7$ used in the model implies that the magnetic flux-tubes whose rise to
the surface produces the $\alpha$-effect are formed at low latitudes
\citep{Kit_20Tube}. The value $B_0 = 10^4$\,G of the $\alpha$-effect
quenching parameter in Eq.\,(\ref{9}) gives reasonable results for the
Sun \citep{Kit_Nep_17AL}. This value was used for the star of one solar
mass. For other masses, the parameter was re-scaled in proportion to the
square root of density at the inner boundary, $B_0 = 10^4\sqrt{\rho_{\rm
i}(M)/\rho{\rm_i}(1M_\odot)}$\,G, to reflect the scaling of flux-tube
rise velocity with the Alfven velocity \citep{D'Silva_Choudhuri_93}.

Diffusive part of the EMF of our model reads
\begin{equation}
    {\vec{\cal E}}^{\rm diff} = -\eta{\vec\nabla}\times{\vec B}
    - \eta_\|{\hat{\vec\Omega}}\times\left({\hat{\vec\Omega}}\cdot{\vec\nabla}\right)
    {\vec B} ,
    \label{11}
\end{equation}
where ${\hat{\vec\Omega}} = {\vec\Omega}/\Omega$ is the unit vector
along the rotation axis. Magnetic diffusivity of Eq.\,(\ref{11}) is
anisotropic: the diffusivity $\eta$ for a direction normal to the
rotation axis is smaller than the diffusivity $\eta + \eta_\|$ along
this axis. The anisotropy is caused by rotation,
\begin{equation}
    \eta = \eta_{_\mathrm{T}}\phi(\Omega^*), \ \ \eta_\| = \eta_{_\mathrm{T}}\phi_\|(\Omega^*).
    \label{12}
\end{equation}
The functions $\phi(\Omega^*)$ and $\phi_\|(\Omega^*)$ of the Coriolis number
\begin{equation}
    \Omega^* = 2\tau\Omega
    \label{13}
\end{equation}
are given in \citet{Kit_Pip_Rue_94}. The rotationally induced anisotropy
is important for the differential rotation model.
Only with account for anisotropy of the eddy heat transport, can the helioseismological
rotation law be reproduced \citep{Rue_Kit_Holl_13}. We include the
diffusion anisotropy in the dynamo model for consistency and for its
better performance \citep{Pip_Sokol_Usoskin_12}.

The diamagnetic pumping results from inhomogeneity of the turbulence intensity \citep{Krause_Raedler_80}. Our dynamo model employs the anisotropic pumping effect for rotating fluids as it was derived by \citet{Kit_Nep_16Dia},
\begin{equation}
    {\vec{\cal E}}^{\rm dia} = -({\vec\nabla}{\tilde\eta})\times{\vec B}
    + ({\vec\nabla}\eta_\|)\times{\hat{\vec\Omega}}({\hat{\vec\Omega}}\cdot{\vec B})\,,
    \label{14}
\end{equation}
with another diffusivity coefficient
\begin{equation}
    {\tilde\eta} = \eta_{_\mathrm{T}}\phi_1(\Omega^*) .
    \label{15}
\end{equation}
Allowance for the pumping effect generally improves performance of the solar dynamo
models \citep{Guerrero_GDP_08,Karak_Cameron_16,Zhang_Jiang_22}.

Magnetic eddy diffusivity $\eta_{_\mathrm{T}}$ can be estimated using
the computed eddy viscosity of Eq.\,(\ref{3}), $\eta_{_\mathrm{T}} =
\nu_{_\mathrm{T}}/{\rm Pm}$, where Pm is the magnetic Prandtl number.
The problem however is that our differential rotation model uses local
mixing-length approximation and does not include the low diffusivity
layer of overshoot convection that is important for the dynamo. We
reduce the diffusivity near the inner boundary to model the layer:
\begin{equation}
    \eta_{_\mathrm{T}} = \frac{1}{\mathrm{Pm}}\left[ \nu_\mathrm{i} +
    \frac{1}{2}(\nu_{_\mathrm{T}} - \nu_\mathrm{i})
    \left(1 + \mathrm{erf}\left(\frac{r/r_\mathrm{i} - x_\eta}{d}\right)\right)\right] ,
    \label{16}
\end{equation}
where $\nu_{\rm i} = 10^{-4}\nu_{\rm max}$ ($\nu_{\rm max}$ is the maximum value of $\nu_{_\mathrm{T}}$ within the convection zone, Pm = 3 in all computations of this paper.
Parameters in the Eqs.\,(\ref{2}) and (\ref{16}) for the star of $1M_\odot$ are taken to be $x_\ell = 1.01,\  d = 0.025,\ x_\eta = 1.1$.

\begin{figure}
\centering
\includegraphics[width= 12 truecm]{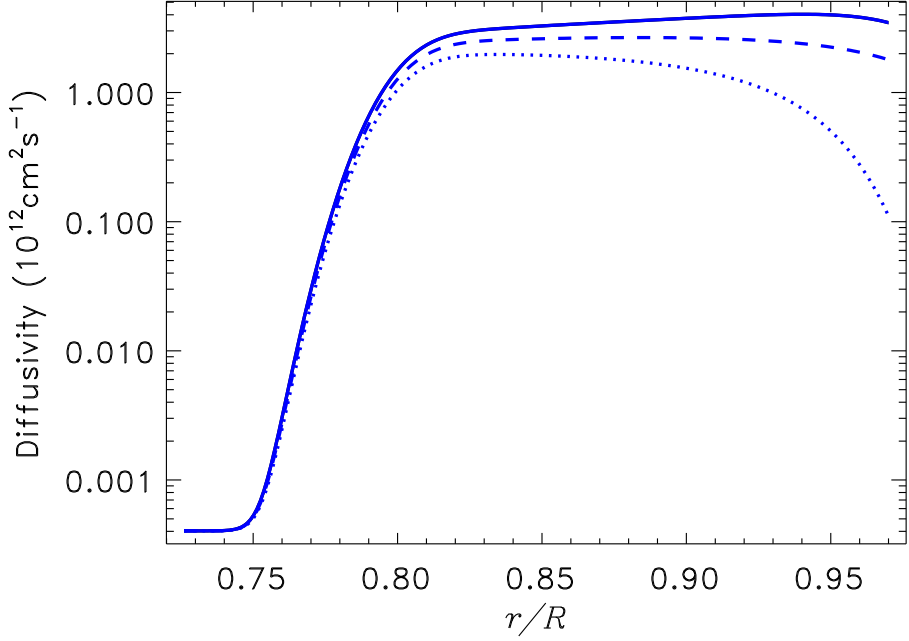}
   \caption{Depth profiles of the diffusion coefficients $\eta$ (full line),
            $\eta_\|$ (dotted) and $\tilde\eta$ (dashed) for a star of $1M_\odot$
            rotating with a period of 23.5 days.}
\label{f2}
\end{figure}

Figure~\ref{f2} shows depth profiles of the diffusivity coefficients of
Eqs.\,(\ref{11}) and (\ref{15}) for the $1M_\odot$ star. The large
\citep[but realistic,][]{Cameron_Schussler_16Eta} diffusivity of this
Figure can result in too short dynamo-cycles. Cycle periods of about
10\,yr in our model are due to the downward diamagnetic pumping that
transports the magnetic field into the near-bottom layer of low
diffusion.

The $\eta_\|$ coefficient decreases towards the surface in
Fig.\,\ref{f2}. This is because the depth-dependent Coriolis number of
Eq.\,(\ref{13}) decreases towards the surface leading to a smaller
rotation-induces diffusion anisotropy.

The parameter values $x_\ell = 1.01,\  d = 0.025,\ x_\eta = 1.1$ were
formerly applied to stars of various mass \citep{Kit_Nep_17}. It has
been realized since then that the near bottom layer of small diffusion
occupies a larger part of the convection zone in stars of larger mass in
this case (a half of the thin convection zone of the $1.2M_\odot$ star).
To avoid such a non-physical prescription, we re-scale the parameters so
that all the characteristic scales constitute the same fractions of the
convection zone thickness in stars of different mass:
\begin{equation}
    d = 0.0745\left(\frac{r_{\rm e}}{r_{\rm i}} - 1\right),\ \
    x_\ell = 0.03\frac{r_{\rm e}}{r_{\rm i}} + 0.97,\ \
    x_\eta = 0.3\frac{r_{\rm e}}{r_{\rm i}} + 0.7 .
    \label{17}
\end{equation}
With this prescription, Fig.\,\ref{f2} looks almost the same for stars
of all considered mass except for the varying minimum value of $r/R$ in
the plot. Some difference in the results of this paper with
\citet{Kit_Nep_17} is explained by the different prescription for the
parameters of the Eq.(\ref{17}) and in the definition of $P_{\rm max}$
by Eq.\,(\ref{1}) which was not yet aware of in 2017.

The dynamo model solves numerically the initial value problem for dynamo
equations (\ref{7}) with a perfect conductor boundary condition imposed
on the bottom and vertical field condition on the top. The initial
condition prescribes the zero toroidal field and the potential
\begin{equation}
    A_0(r,\theta) = \frac{B_N(r - r_{\rm i})(2r_{\rm e} - r_{\rm i} - r)}
    {4(1 - r_{\rm i}/r_{\rm e})^2}
    [1 - p + (1 + p)\cos\theta]\sin^2\theta
    \label{18}
\end{equation}
for the poloidal field, where $B_N$ is the field strength on the
northern pole and $-1 \leq p \leq 1$ is the parity index ($p = 1$ means
a quadrupolar equator-symmetric initial field and it is $p = -1$ for a
dipolar antisymmetric field). Starting from the initial condition of
mixed parity, the dynamo code was run for one thousand years simulated
time. This preliminary run suffices for the dynamo to converge to a
periodic oscillation, for which the cycle period and other results of
Sect.\,\ref{4} were obtained. If the initial condition (\ref{18}) had a
certain parity ($p = \pm 1$), the numerical solution cannot depart from
this parity. The runs with so-prescribed parity helped to compute the
threshold amplitude $\alpha_{\rm c}$ of the $\alpha$-effect of
Eq.\,(\ref{9}) for the onset of dynamo-instability for dipolar
($\alpha_{\rm c}^{\rm d}$) and quadrupolar ($\alpha_{\rm c}^{\rm q}$)
fields.
%%%%%%%%%%%%%%%%%%%%%%%%%%%%%%%%%%%%%%%%%%%%%%%%%%%%%%%%%%%%%%%%%%%%%%%%%
\subsection{Estimating rotation period and structure of stars}\label{S3.2}
%%%%%%%%%%%%%%%%%%%%%%%%%%%%%%%%%%%%%%%%%%%%%%%%%%%%%%%%%%%%%%%%%%%%%%%%%
Computations of differential rotation and dynamo require the structure
and rotation rate of a star to be specified.

We use the gyrochronology relation by \citet{Barnes_07}
\begin{equation}
     P_\mathrm{rot} = a t^n\left(B - V\ -\ 0.4\right)^b\ \mathrm{d},
    \label{19}
\end{equation}
relating the age $t$ (in Myr) and $B-V$ colour of a star to the rotation
period. The parameter values of $a = 0.77,\ n = 0.512,\ b = 0.6$ within
their uncertainty range reproduce closely the case with the Sun.

\begin{figure}
\centering
\includegraphics[width= 12 truecm]{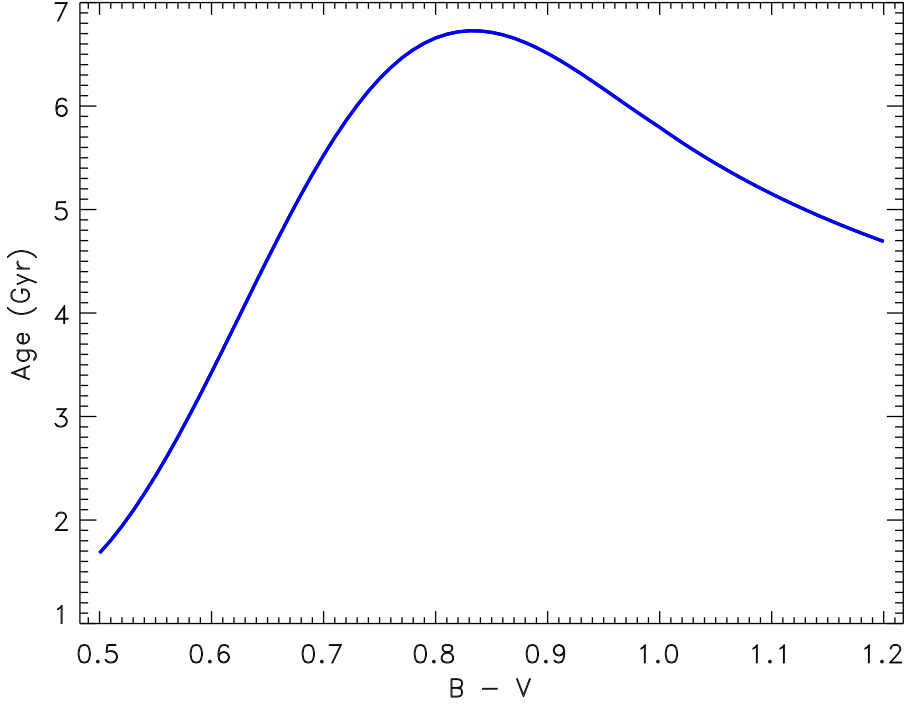}
   \caption{Stellar age of the maximum rotation period $P_{\rm max}$ as the function of $B-V$ colour estimated with Eq.\,(\ref{20}).}
\label{f3}
\end{figure}

Gyrochronology is believed to overestimate the rotation period after the
rotation slows down to the minimum rate of Eq.\,(\ref{1})
\citep{van_Saders_EA_16,van_Saders_EA_19}. Stellar structure varies
slowly at the age when it happens. We therefore assume that the relation
(\ref{19}) applies up to the age when the maximum rotation period of
Eq.\,(\ref{1}) is attained. This age can be roughly estimated from the
reversed Eq.\,(\ref{19}):
\begin{equation}
    t = \left[ \frac{2.08\ \tau_{\rm c}}{a (B-V\,-\,0.4)^b} \right]^{1/n} .
    \label{20}
\end{equation}
Figure\,\ref{f3} shows the age of Eq.\,(\ref{20}) as function of $B - V$
colour. This age of the  large-scale dynamo termination does not decrease
monotonously with increasing temperature.

The {\sl EZ} model by \citet{Paxton_04EZ} was used to define the
evolutionary sequence of structure models for a star of given mass and
metallicity $Z = 0.02$. The colour-temperature relation and the
interpolation code by \citet{VandenBerg_Clem_03} was used to estimate
the $B-V$ color corresponding to the structure models. The rotation
period of Eq.(\ref{19}) was then compared with the $P_{\rm max}$ of
Eq.\,(\ref{1}). The structure model with the closest values of these two
rotation periods is assumed to correspond to the star that arrived on
the dashed line of Fig.\,\ref{f1}. The Sun is almost on this line. The
solar dynamo was estimated to be about 10\% supercritical in the sense
of the amplitude $\alpha$ of the $\alpha$-effect of Eq.\,(\ref{9}) \citep{Kit_Nep_17}. The differential rotation of the stars of different mass which \lq arrived
on the dashed line' of Fig.\,\ref{f1} was computed and then used in the
simulations of their 10\% supercritical dynamos as explained at the end
of Sect.\,\ref{S3.1}.

The computations cover the mass range from $0.7M_\odot$ to $1.2M_\odot$
with increment of $0.05M_\odot$.
%%%%%%%%%%%%%%%%%%%%%%%%%%%%%%%%%%%%%%%%%%%%%%%%%%%%%%%%%%%%%%%%%%%%%%%%%
\section{Results and discussion}\label{S4}
\begin{figure}
\centering
\includegraphics[width= 12 truecm]{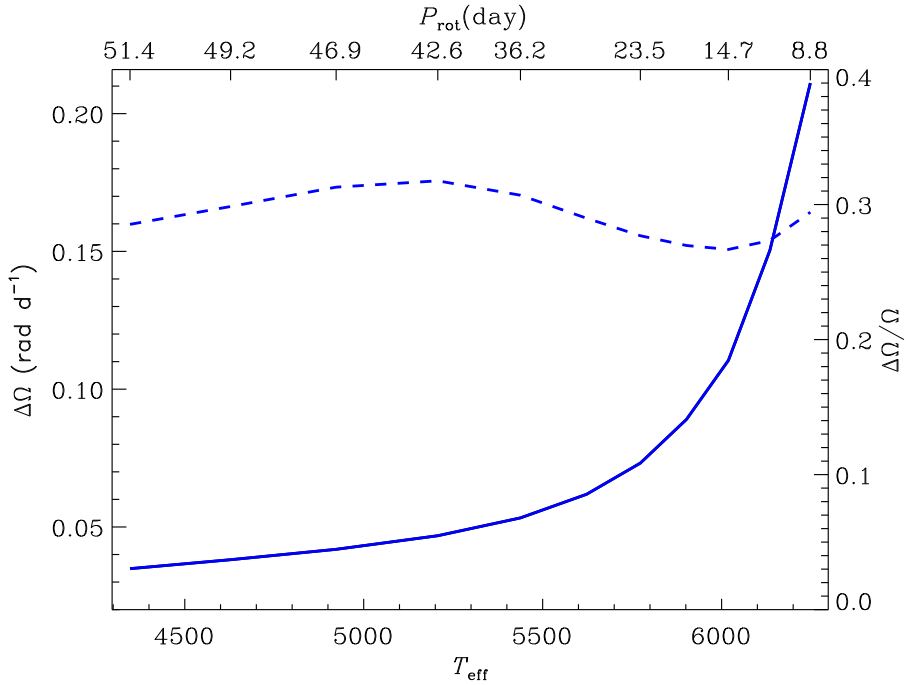}
   \caption{{\sl Full line, left scale:} difference in rotation rates
            between equator and pole as the function of temperature. {\sl Dashed line, right scale:} the relative value $\Delta\Omega / \Omega$ of the differential rotation. The scale on the top shows the rotation period corresponding to the dashed line of
            Fig.\,\ref{f1}.}
\label{f4}
\end{figure}

%%%%%%%%%%%%%%%%%%%%%%%%%%%%%%%%%%%%%%%%%%%%%%%%%%%%%%%%%%%%%%%%%%%%%%%%%
All computations show solar-type differential rotation with faster
equatorial rotation. Figure\,\ref{f4} shows the surface equator-to-pole
difference in rotation rate in dependence on the effective temperature.
Hotter stars have larger differential rotation. As the hotter stars
rotate faster, this Figure also means an increase in the differential
rotation with rotation rate. Figure\,\ref{f4} is a slow rotation
counterpart of the observational fig.\,2 by \citet{Barnes_EA_05} and
theoretical fig.\,10 by \citet{Kit_Ole_11DR}.

Figure\,\ref{f4} also shows the dimensionless ratio $\Delta\Omega /
\Omega$, which varies little with $T_{\rm eff}$. As the computations of
this Figure were done for constant Ro = 2.08, the small variation in
$\Delta\Omega / \Omega$ means that $\Delta\Omega\tau_{\rm c}$ does also
vary little with $T_{\rm eff}$. The increase in $\Delta\Omega$ with
$T_{\rm eff}$ proceeds in inverse proportion to decreasing convective
turnover time. The scaling with $\tau_{\rm c}^{-1}$ can explain the
strong increase in differential rotation with temperature observed by
\citet{Barnes_EA_05}.

\begin{figure}
\centering
\includegraphics[width= 12 truecm]{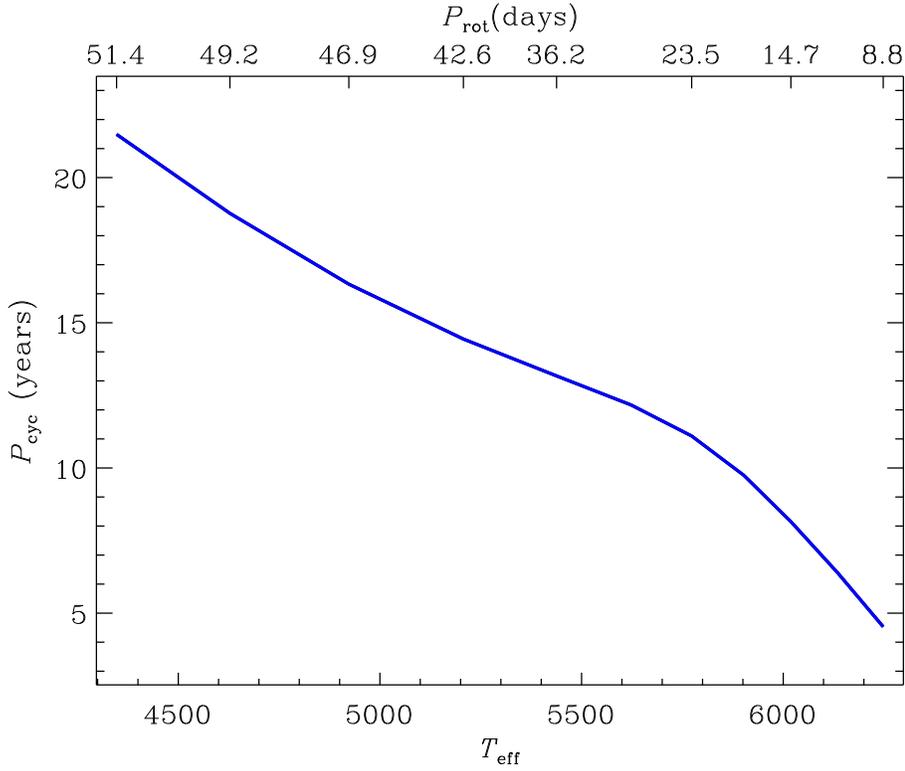}
   \caption{Periods of the computed dynamo-cycles as the function
            of temperature. Similar to Fig.\,\ref{f4}, the scale on the top shows the corresponding rotation period.}
\label{f5}
\end{figure}

Figure\,\ref{f5} shows a similar plot for the period of computed dynamo
cycles. These are the periods of energy oscillation (half-periods of the
sign-changing magnetic cycles). Dynamo computations predict shorter
cycles for hotter stars. For fixed Rossby number of Eq.\,(\ref{1}),
this temperature trend also implies a shorter cycle for faster
rotation. Similar to the differential rotation, the observed decrease in
activity cycle duration with rotation rate can be at least partly
explained by its temperature dependence. In difference with the
differential rotation, we did not find a normalization for $P_{\rm
cyc}$, which varies little with $T_{\rm eff}$.

\begin{figure}
\centering
\includegraphics[width= 12 truecm]{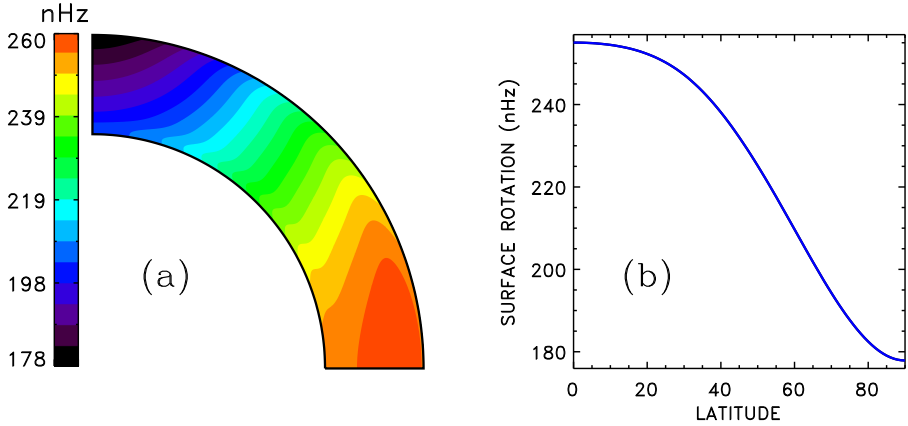}
   \caption{Differential rotation of the $0.8M_\odot$ star. (a) Rotation rate
            isolines in the NW quadrant of the meridional cross-section
            of the convection zone. (b) Latitudinal profile of the
            surface rotation rate.}
\label{f6}
\end{figure}
\begin{figure}
\centering
\includegraphics[width= 12 truecm]{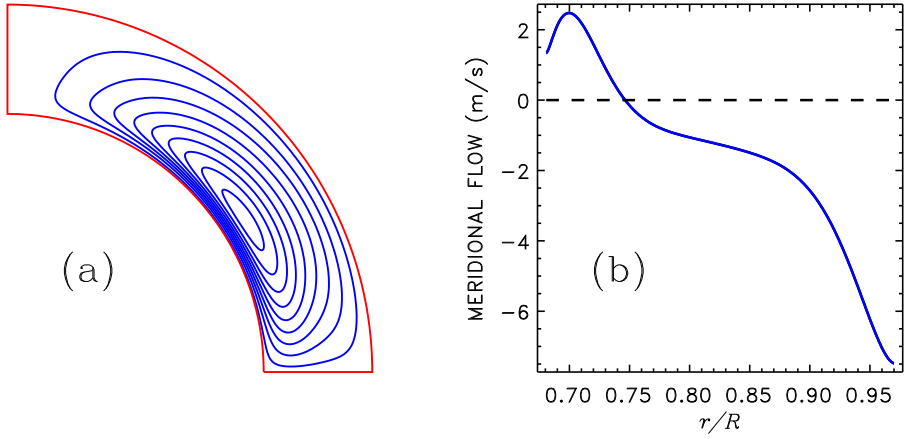}
   \caption{Meridional flow in the $0.8M_\odot$ star. (a) Stream lines of the flow. (b) Variation of the meridional velocity with radius at the 45$^\circ$ latitude. Positive velocity means equator-ward flow.}
\label{f7}
\end{figure}
\begin{figure}
\centering
\includegraphics[width= 12 truecm]{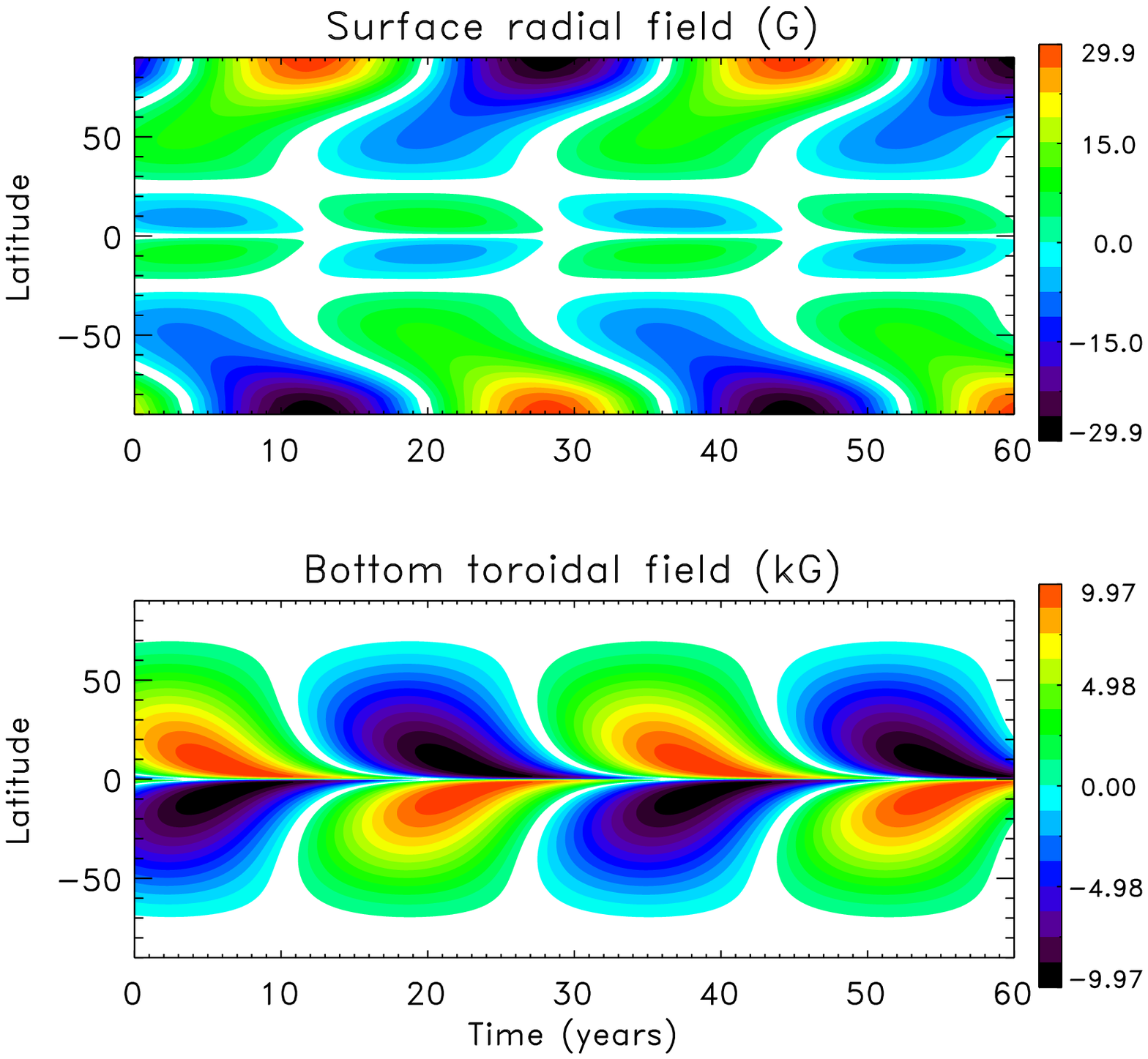}
   \caption{Time-latitude diagrams of the dynamo model for $0.8M_\odot$ star.
            Surface radial field and the bottom toroidal field are shown in the top and bottom panels respectively.}
\label{f8}
\end{figure}

The difference with the differential rotation also is that the
theoretical dependence of cycle period on rotation rate for fixed
temperature is uncertain and may not be weak. Non-kinematic dynamo
models are required to study this dependence. Consideration of
Sect.\,\ref{S2} suggests that activity of rapidly rotating young stars
may not be cyclic. \citet{Katsova_EA_15} estimated that the Sun formed
its activity cycle at the age of 1 to 2 Gyr. This means about two times
faster rotation compared to its present rate. Red symbols in
Fig.\,\ref{f1} are not very close to the dashed line of $P_{\rm max}$.

We consider next the differential rotation and dynamo for two cases of
stellar mass smaller and larger compared to the Sun. The consideration
shows that equatorial symmetry of the dominant dynamo mode does also
depend on temperature.

\begin{figure}
\centering
\includegraphics[width= 12 truecm]{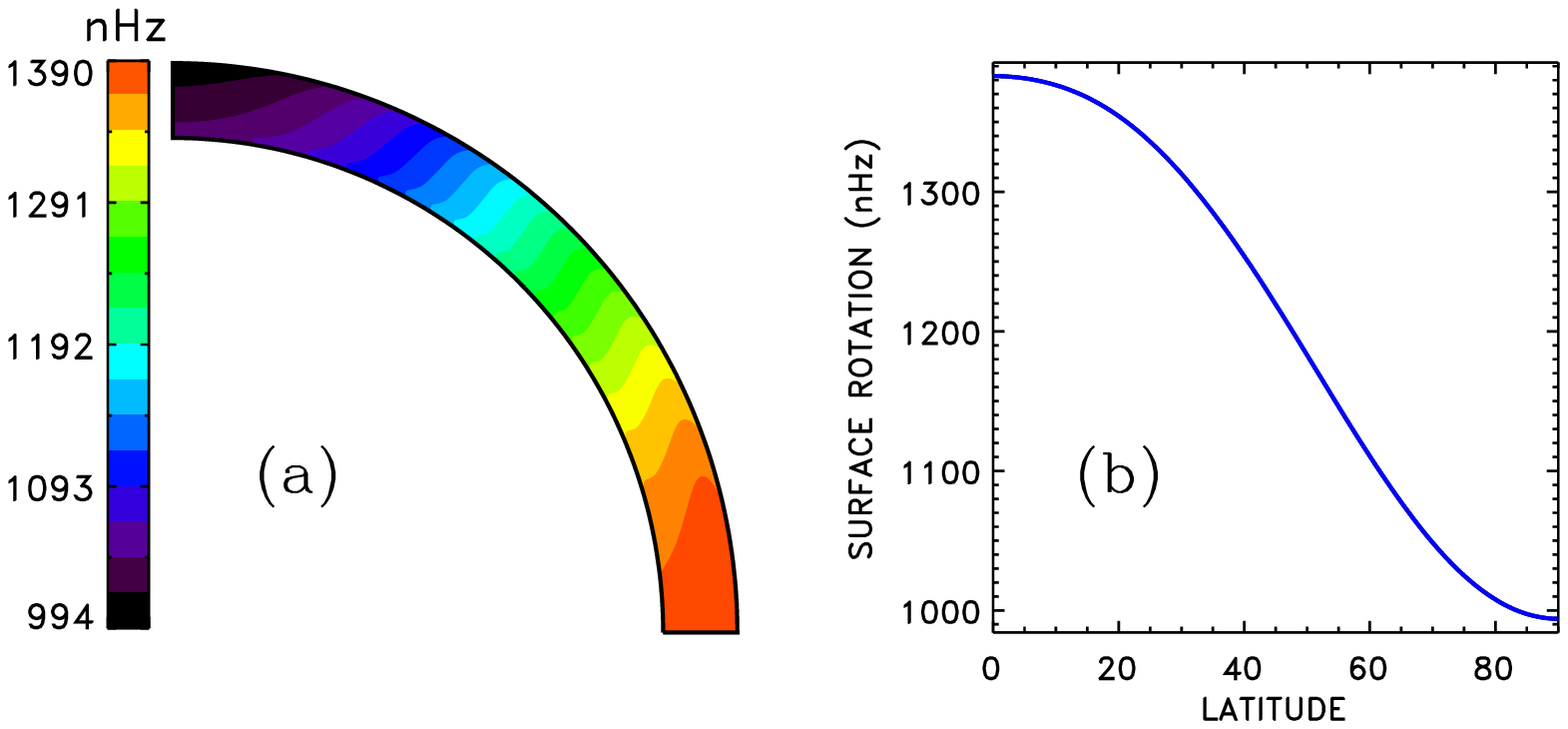}
   \caption{Differential rotation of $1.2M_\odot$ star. (a) Rotation rate isolines.
            (b) Latitudinal profile of the surface rotation rate.}
\label{f9}
\end{figure}
\begin{figure}
\centering
\includegraphics[width= 12 truecm]{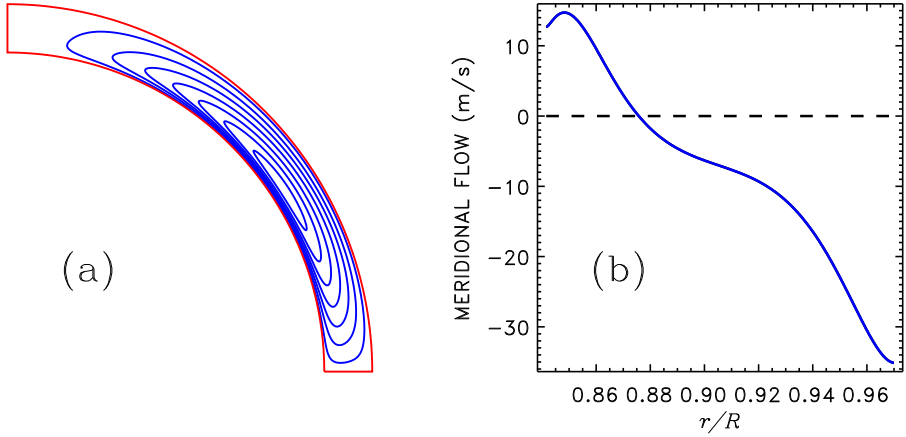}
   \caption{Meridional flow in the $1.2M_\odot$ star. (a) Stream lines of the flow.
            (b) Depth profile of the meridional velocity at the 45$^\circ$ latitude. Positive velocity means equator-ward flow.}
\label{f10}
\end{figure}
\begin{figure}
\centering
\includegraphics[width= 12 truecm]{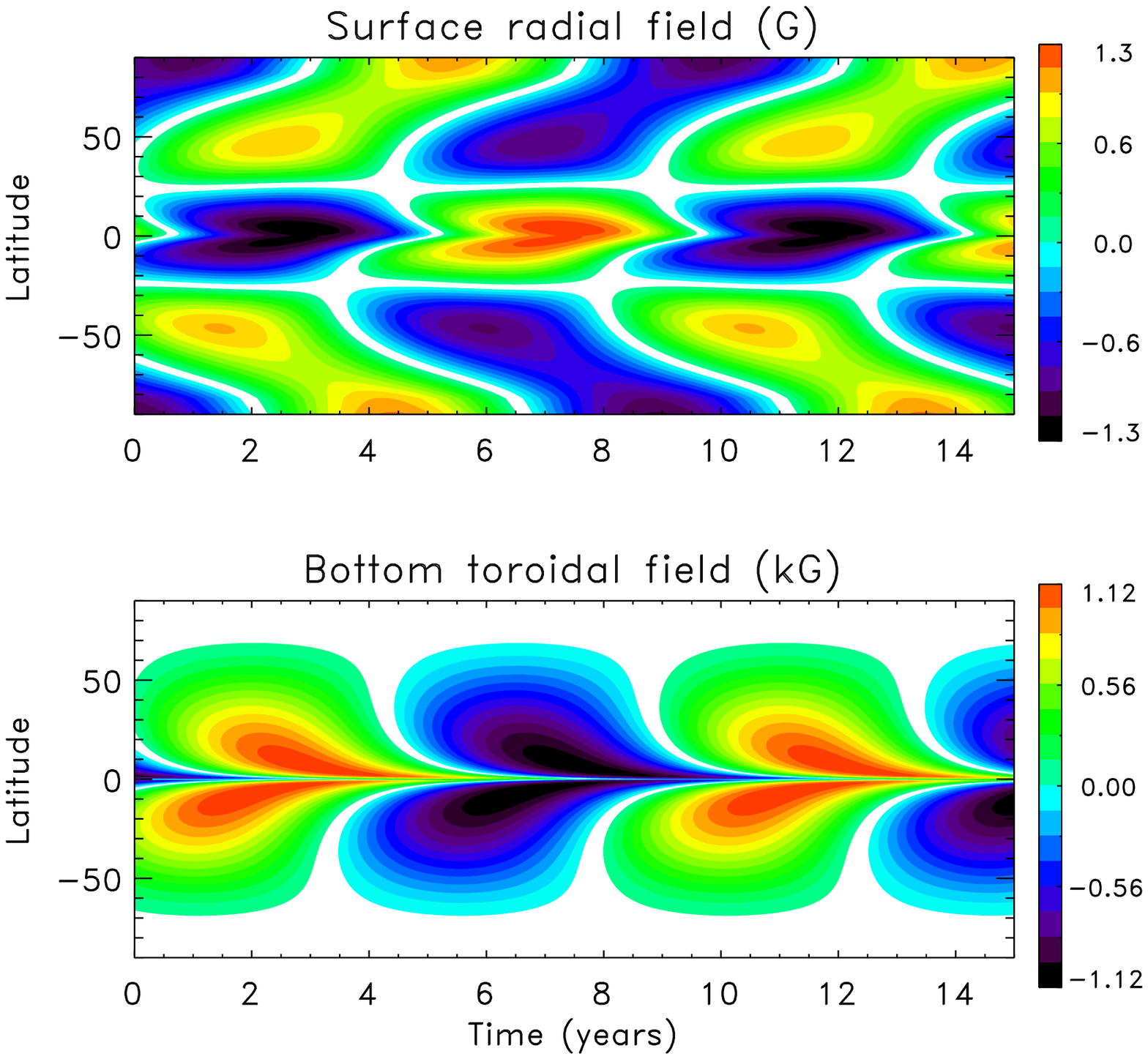}
   \caption{Time-latitude diagrams of the dynamo model for $1.2M_\odot$ star.
            Surface radial field and the bottom toroidal field are shown in the top and bottom panels respectively.}
\label{f11}
\end{figure}

Figures \ref{f6} to \ref{f8} show the differential rotation, meridional
flow and magnetic time-latitude diagram computed for the $0.8M_\odot$
star. Similar to the Sun, dipolar dynamo mode dominates in this case.
The dynamo arrived at dipolar parity from a mixed-parity initial state
of Eq.\,(\ref{18}).

\begin{figure}
\centering
\includegraphics[width= 12 truecm]{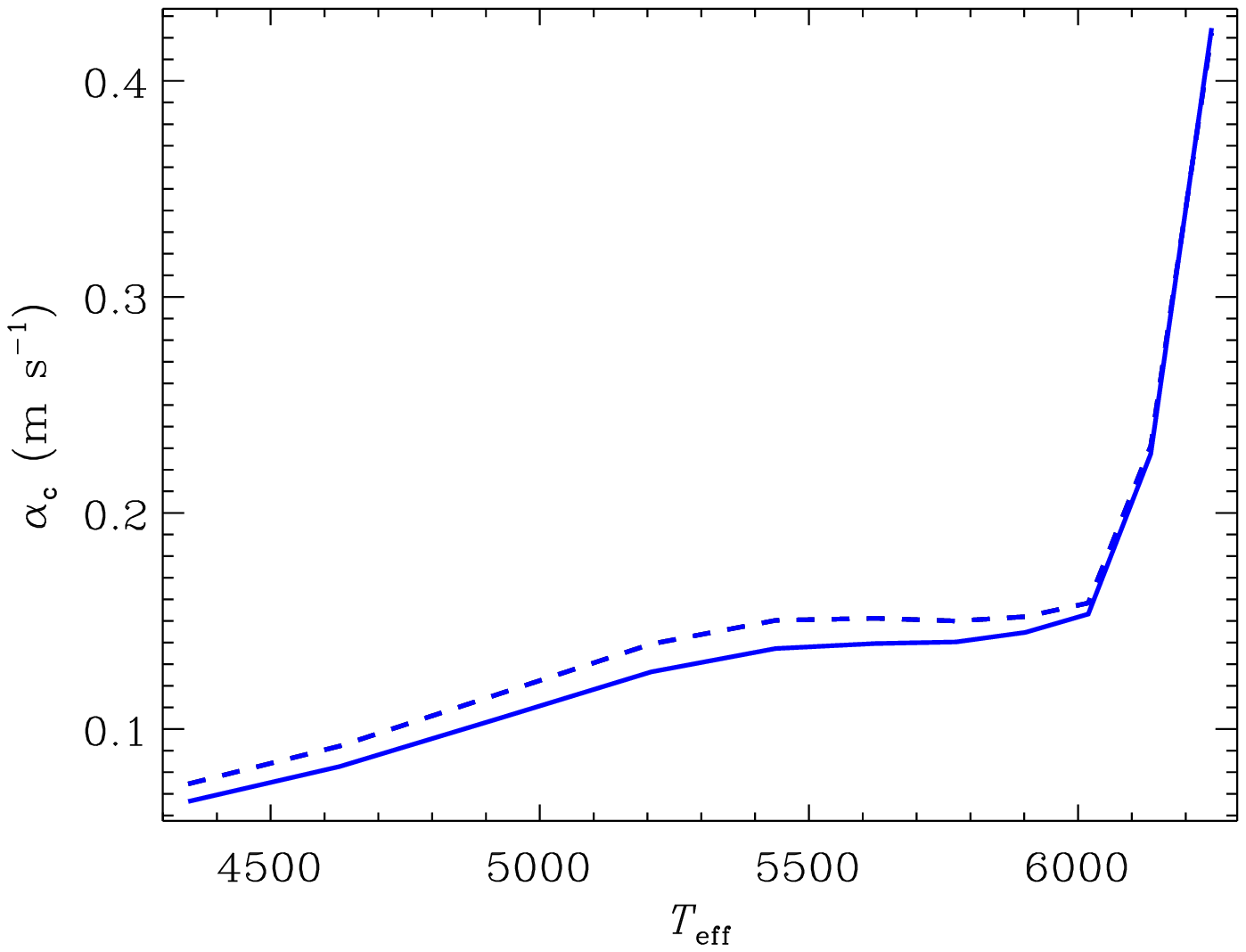}
   \caption{Marginal amplitude $\alpha_{\rm c}$ of the $\alpha$-effect for generation of dipolar (full line) and quadrupolar (dashed) magnetic fields as the function of temperature.}
\label{f12}
\end{figure}

Figures \ref{f9} to \ref{f11} show the differential rotation, meridional flow and field
diagram computed for the $1.2M_\odot$ star. The differential rotation pattern of Fig.\,\ref{f9} is similar to Fig.\,\ref{f6} in spite of about five times faster rotation of the $1.2M_\odot$ star. This result contrasts with the dependence of differential rotation on rotation rate produced by our model for fixed stellar mass. The differential rotation changes towards a cylinder-shaped pattern with increasing rotation in a star of given mass. The change results in a weaker meridional flow for faster rotation
\citep[see figs.\,4 and 5 in][]{Kit_Ole_12DR}. The weakening of meridional flow is the probable reason for longer activity cycles in faster rotating stars (of a given mass) found by \citet{Brun_Cyc_10} and \citet{Karak_EA_14} with the flux-transport dynamo models. A change towards cylinder-shaped rotation does not happen between Figs.\,\ref{f6} and \ref{f9} of this paper because these Figures correspond to computations for different stellar mass but the same Rossby number of Eq.\,(\ref{1}). The effect of rotation on large-scale flow is measured by the dimensionless Coriolis number $\Omega^* = 4\pi {\rm Ro}^{-1}$ of Eq.\,(\ref{13}). The characteristic Coriolis number has the same value of $\Omega^* \simeq 6$ in all computations of this paper. This value is below the range of $10 < \Omega^* < 20$ where the change to cylinder-shaped rotation occurs \citep{Kit_Ole_12DR}. Deviation from cylinder-shaped rotation is caused by a slight increase of the mean temperature with latitude in the convection zone. The differential temperature results in our model from the rotationally induced anisotropy of the eddy heat transport \citep{Ruediger_EA_05Heat_flux}. The anisotropy is controlled by the Coriolis number of Eq.\,(\ref{13}). Computations for different stellar mass and rotation rate but the same characteristic value of $\Omega^*$ give similarly shaped differential rotation.

Computations for the $1.2M_\odot$ star show
dynamo convergence to the mixed parity solution.
The field diagram of Fig.\,\ref{f11} shows mixed equatorial symmetry.
Convergence to a certain symmetry requires a certain link between the
northern and southern hemispheres. The hemispheric link weakens with
decreasing thickness of the convection zone in stars of larger mass.

Another manifestation of the weaker hemispheric link in the thin
convection zones is the almost equal
threshold amplitudes $\alpha_{\rm c}^{\rm d}$ and $\alpha_{\rm c}^{\rm
q}$ for generation of dipolar and quadrupolar fields in Fig.\,\ref{f12}.
The $\alpha_{\rm c}$ of this Figure increases with temperature. A larger
$\alpha_{\rm c}$ needs faster rotation. This can explain the shorter
$P_{\rm max}$ for more massive stars (smaller $B - V$) in Fig.\,\ref{1}.
The spindown is caused by the large-scale magnetic fields increasing the co-rotation radius for the stellar wind plasma.
The spindown stops when rotation slows to relatively large rate corresponding to the relatively large $\alpha_{\rm c}$ in more massive stars.
Further increase in $\alpha_{\rm c}$ for $M > 1.2M_\odot$ may eventually
mean that even $P_{\rm rot} \sim 1\,{\rm d}$ of {\sl ZAMS} stars does
not suffice for a dynamo. Based on observations,
\citet{Durney_Latour_78F4} concluded that stars of spectral type earlier
than F6 do not support large-scale dynamos.

The $\alpha$-effect in our computations is 10\% supercritical, $\alpha =
1.1\,\alpha_{\rm c}$. Computations with other slightly supercritical
$\alpha$ show that the amplitude $B$ of magnetic cycles is proportional
to the square root of the super-criticality, $B \propto (\alpha -
\alpha_{\rm c})^{1/2}$. This relation holds not only for the
dynamo-instability, but is a general rule for any weakly
nonlinear instability \citep[cf. eq.\,(26.10)
in][]{Landafshitz_Hydro_87}. The relation can be
reformulated in terms of the rotation rate,
\begin{equation}
    B \propto \left(P_{\rm max} - P_{\rm rot}\right)^{1/2} ,
    \label{21}
\end{equation}
and possibly explain why the Sun is observed to be less active than
other stars of comparable effective temperature and rotation rate
\citep{Inactive_Sun_20Sci,Inactive_Sun_20}. According to
Eq.\,(\ref{21}), what matters for magnetic activity is not the value of
the rotation rate but the amount of excess by the rate its marginal
value for dynamo. The derivative of Eq.\,(\ref{21}) on $P_{\rm rot}$ is
infinite at $P_{\rm rot} = P_{\rm max}$. Stars with close rotation rates
can differ considerably in super-criticality of their dynamos and
therefore in the level of magnetic activity.

The ratio of the cycle period to the time of advection by the meridional
flow, $P_{\rm cyc}V_{\rm bot}/r_{\rm i}$, varies little to remain
between the values of 3 and 4 in our computations; $V_{\rm bot}$ is the
near-bottom maximum value of the meridional flow velocity (see Figs
\ref{f7}b and \ref{f10}b). This means that the computations belong to
the flux-transport dynamo regime. Faster meridional flow in hotter
stars explain their shorter cycles.

A preliminary run of one thousand years starting from the mixed-parity
initial state of Eq.\,(\ref{18}) did not converge to a certain
equatorial symmetry for $1.2M_\odot$ star though $\alpha_{\rm c}^{\rm
q}$ is slightly smaller than $\alpha_{\rm c}^{\rm  d}$ in this case
(Fig.\,\ref{f11}). We did not extend the run further for the following
reason. Our computations do not include fluctuations in dynamo
parameters, which are most probably present in stars. The
equator-asymmetric fluctuations couple the dipolar and quadrupolar
dynamo modes so that the amplitudes of these modes vary irregularly on a
time scale comparable to the cycle period
\citep{Cameron_Schussler_18NSAsymm,Kit_Khlystova_21}. The amplitudes are
expected to be comparable for almost equal super-criticality of the two
modes. This will result in irregularly varying north-south asymmetry in
activity of a star. Observational detection of an activity cycle may be
difficult in this case if the inclination angle of the rotation axis is
not close to $\pi/2$. This may be the reason for no detections of
high-quality cycles for stars hotter than the Sun (Fig.\,\ref{f1}).
%%%%%%%%%%%%%%%%%%%%%%%%%%%%%%%%%%%%%%%%%%%%%%%%%%%%%%%%%%%%%%%%%%%%%%%%%
\section{Conclusions}\label{S5}
%%%%%%%%%%%%%%%%%%%%%%%%%%%%%%%%%%%%%%%%%%%%%%%%%%%%%%%%%%%%%%%%%%%%%%%%%
Young rapidly rotating stars are not probable to show sun-like activity
cycles. Otherwise, the large-scale stellar dynamos would be an exception
among other hydromagnetic instabilities showing turbulence rather than
cyclic overstability in a highly supercritical regime.

The upper bound on the rotation period of main-sequence dwarfs showing
solar-type magnetic activity
\citep{Rengarajan_84,Metcalfe_EA_16,van_Saders_EA_16} can be interpreted
as the minimum rotation rate for a large-scale stellar dynamos. This
rotation rate increases with the effective temperature. Computations
with a joint model for differential rotation and dynamo show magnetic
cycles for slightly supercritical dynamos in stars of different mass.
Hotter stars have shorter cycles in the computations and the hotter
stars rotate faster. The observed decrease in cycle duration for faster
rotation based on combined statistics of stars of different spectral
types can, therefore, be at least partly explained by the cycle
dependence on temperature.

The computations also show larger marginal values of the $\alpha$-effect
for dynamo operation in hotter stars. A larger $\alpha$ demands faster
rotation. This may be the reason for the smaller upper bound on the
rotation period for hotter stars. The amplitude of magnetic energy in
the dynamo model is proportional to the difference between rotation rate
and the marginal rate for dynamo (cf. Eq.\,\ref{21}). Stars with similar
rotation rates can therefore differ substantially in level of their
activity as observed \citep{Inactive_Sun_20Sci}: a small difference in
rotation rates does not necessarily mean an equally small difference in the
super-criticality.

Large variations in differential rotation and cycle period computed
for the constant Rossby number of Eq.\,(\ref{1}) indicate that this
number may not be the universal scaling parameter for stellar rotation
and dynamos. The Rossby number measures intensity of interaction between
convection and rotation. It can be doubted that the BL-mechanism of the
solar-type dynamos is fully controlled by this interaction.

The dynamo computations predict a change in equatorial symmetry of
global magnetic fields with temperature. The stars of solar and smaller
mass show antisymmetric fields about the equator in the computations. A
change to mixed-parity asymmetric fields is predicted for more massive
stars. The mixed-parity dynamos can impede observational detection of
the activity cycles.

The well-defined activity cycle of subgiant HD\,81809 is very
interesting and challenging to dynamo theory (see the footnote\,1). This
star exceeds the Sun in mass $M = (1.70 \pm 0.64)M_\odot$ \citep[see
table 2 in][]{Egeland_18}. A main-sequence A-star is its probable
progenitor. A-stars do not have (sufficiently thick) external convection
zones and do not show activity cycles. The convection zone formed when
HD\,81809 evolved from the main-sequence is probably responsible for its
cyclic activity. This example can be informative on the role of
convective envelopes for stellar dynamos.
%%%%%%%%%%%%%%%%%%%%%%%%%%%%%%%%%%%%%%%%%%%%%%%%%%%%%%%%%%%%%%%%%%%%%%%%%
\begin{acknowledgement}
The author is thankful to an anonymous referee for pertinent and
constructive comments and to Maria Katsova for a useful discussion. This
work was financially supported by the Ministry of Science and High
Education of the Russian Federation.
\end{acknowledgement}
%%%%%%%%%%%%%%%%%%%%%%%%%%%%%%%%%%%%%%%%%%%%%%%%%%%%%%%%%%%%%%%%%%%%%%%%%
\bibliographystyle{raa}
\bibliography{paper}
%%%%%%%%%%%%%%%%%%%%%%%%%%%%%%%%%%%%%%%%%%%%%%%%%%%%%%%%%%%%%%%%%%%%%%%%%
\label{lastpage}
%%%%%%%%%%%%%%%%%%%%%%%%%%%%%%%%%%%%%%%%%%%%%%%%%%%%%%%%%%%%%%%%%%%%%%%%%
\end{document}